# Effects of Local Concentration Gradients of Monocyte Chemoattractant Protein-1 (MCP-1) on Monocytes Adhesion and Transendothelial Migration in a Three-Dimensional (3D) *In Vitro* Vascular Tissue Model

Neda Ghousifam[1*], Mohamad Eftekharjoo[2], Tahereh Derakhshan[1],

and Heather Gappa-Fahlenkamp[1]

*Corresponding author

[1]School of Chemical Engineering, Oklahoma State University, 420 Engineering North,

Stillwater, OK 74078, USA

[2]Department of Mechanical and Aerospace Engineering, Old Dominion University,

Norfolk, VA, 23529, USA

E-mail: *neda.ghousifam@okstate.edu

Phone: *1-405-762-3603

**Abstract**

*Objective* The main objective of this study was to investigate the formation of MCP-1 concentration gradients within the subendothelial ECM and determine the effect on monocytes migration in response to inflammation. To meet this objective, monocytes migration in a 3D vascular tissue model, containing a matrix where concentration gradients may form, was compared to that in a 2D cell culture model results.

*Materials* The 3D vascular tissue model consists of human aortic endothelial cells (HAEC) grown on the surface of a collagen matrix. The HAEC form an endothelium and the collagen is used to mimic the subendothelial ECM. The 2D cell culture model consisted of HAEC grown on a porous membrane of a Transwell culture insert.

*Results* An overall greater monocytes adhesion and transendothelial migration was observed in the 3D model compared to the 2D model after 24 h stimulation.

*Conclusion* MCP-1 transport is different in the 3D vascular tissue model than the 2D microporous membrane model, which results in a difference in monocytes transendothelial migration between the two models. This research will provide new information about the relationship between the MCP-1 concentration gradient and monocytes transendothelial migration.

# 1. Introduction

Atherosclerosis, the primary cause of cardiovascular disease (CVD) [1], is an inflammatory disease and is characterized by endothelial dysfunction, inflammation, and extracellular matrix (ECM) remodeling [2, 3]. Atherosclerotic plaque formation begins with accumulation of lipids in the subendothelial layer of major arteries, followed by the adhesion of monocytes and lymphocytes to the endothelium, and the subsequent recruitment of these cells across the endothelium and into the ECM [4-9]. Plaque formation can significantly reduce blood flow in an artery, which may cause a stroke or heart attack. The adhesion and transmigration of leukocytes across the endothelium include their rolling, adhesion, and crawling which is followed by their paracellular and transcellular migration through the basement membrane [10, 11]. Generation of inflammatory signals regulates cellular movement and differentiation of leukocytes from the circulation into the injured tissue in order to aid in wound healing processes. Cellular adhesion molecules (CAMs) and inflammatory chemotactic cytokines (chemokines) are responsible for the recruitment of leukocytes to the site of inflammation. Many cardiovascular diseases, such as atherosclerosis, induce inflammation and result in the over-recruitment of leukocytes [9, 12-15]. Past studies show that specific chemokine profiles dictate the migration of leukocytes to the site of inflamed tissue [15, 16].

CAMs are proteins expressed on the surface of the cells and are involved in the adhesion of monocytes to the endothelial cell layer [2, 7, 8, 17-21]. The CAMs that are relevant to vascular diseases include vascular cell adhesion molecule-1 (VCAM-1), intercellular adhesion molecule-1 (ICAM-1), and platelet endothelial adhesion molecule-1 (PECAM-1). Chemokines are low molecular weight proteins (8-10 kDa) known to have a pivotal role in immune cells trafficking and activation [14, 15, 22-25]. The migration of leukocytes from the blood to the tissue site of inflammation is dependent on chemotactic gradients [26-28], such as monocyte chemoattractant protein-1 (MCP-1). MCP-1 (chemotactic toward monocytes and T cells) [13] plays a role in monocytes trafficking across the endothelial layer in early stages of atherosclerosis [12, 29, 30]. Furthermore, it has been shown that MCP-1 expression is highly upregulated in an atherosclerotic lesion [12, 31-34], and the formation of the lesion decreased significantly in the murine models lacking MCP-1 [35, 36].

Previous *in vitro* studies illustrate that the transendothelial migration of monocytes depends on the concentration gradient of free MCP-1 across the endothelial layer [10, 16, 37-42]. The highest concentration of MCP-1 is at the source of atherosclerotic inflammation in the arterial walls [14]. Further, MCP-1 is shown to be secreted from the endothelial cell layer in a free form [43], and when the endothelial cells are stimulated, the secretion of MCP-1 was found to be non-polarized [16]. Based on these two findings, it was suggested that *in vivo*, MCP-1 secreted in the luminal side could be prevented by the blood flow into the vascular lumen, while the concentration gradient of MCP-1 is formed within the ECM via the diffusion of MCP-1 released from the basal side of the endothelial cell layer into the subendothelial ECM [43]. Therefore, formation of the MCP-1 concentration gradient is driving monocytes migration into the ECM. However, in response to an inflammatory stimuli, influx of monocytes to the site of injury increases dramatically, and an atherosclerosis lesion may develop, which is the basis of pathology in atherosclerosis. Presently, profiling the development of MCP-1 gradients in the ECM is still lacking; due to there are no available techniques that can be used to quantify such gradients.

Two-dimensional (2D) cell culture models have been used to investigate the effect of the concentration gradient of free MCP-1 on monocytes migration [37, 41, 42]. The model is adequate for showing a response of monocytes to a chemokine that has formed a concentration gradient across a membrane with/without the endothelial cell layer; however, the 2D model cannot be used to examine chemokine concentration gradients that are present within the tissue *in vivo* [15, 23, 24, 44-52]. In the 2D experimental model, endogenous MCP-1 released by the cells to the surrounding liquid media is diluted, quenching the chemoattractant effect. Hence, the model has been used to investigate monocytes migration in response to free exogenous chemokines. Using 2D models for chemotaxis studies can be too simplistic and overlook what occurs *in vivo* due to the lack of a third dimension [53] and they may not be suitable for predicting what occurs in a complex three-dimensional (3D) model, like those within the human body [53]. A better alternative experimental model is a 3D model consisting of a collagen matrix to mimic the subendothelial ECM [54]. In this study, an advanced 3D *in vitro* vascular tissue model (defined as 3D tissue model throughout the text) was introduced as a tool to study the underlying mechanisms occurring within the ECM in atherosclerosis. The major advantage of the 3D tissue model is that the third dimension provides the supplementary space that is significant for the creation of free and bound MCP-1 diffusive concentration gradients in the ECM [54], which is responsible for the control of many cellular mechanisms.

In our previous studies, we have demonstrated the existence of bound MCP-1 and formation of diffusive concentration gradient in a simplified cell-free 3D matrix and have derived a mathematical model to describe such a gradient [54, 55]. In this study, human aortic endothelial cells (HAEC) are added to the model to form an endothelium on the surface of a collagen matrix used to represent the ECM. MCP-1 released from the endothelial cell layer can diffuse into the ECM and form localized concentration gradients which can influence the monocytes migration. The goal of the current study is to investigate the effect of MCP-1 local concentration gradients on monocytes migration in the 3D tissue model and compare the outcomes to the existing 2D model results. We hypothesize that the MCP-1 concentration gradient within the ECM of the 3D tissue model would drive a different monocytes migration response than in the 2D model based on the difference in the nature of existing concentration gradients in the 3D and 2D microenvironments.

HAEC growth characteristics (i.e., viability, trans-endothelial electric resistance, and morphology), the kinetic and magnitude expression of the CAMs (VCAM-1, ICAM-1, PECAM-1) and MCP-1, and transport properties (MCP-1 release profile) is also studied between the 2D and the 3D models. The experiments are performed under quiescent conditions and when models are immunologically activated with tumor necrosis factor-α (TNF-α). TNF-α was used to mimic inflammation that occurs in the early stages of atherosclerosis. Differences between two models characteristics, such as endothelium cell behavior, MCP-1 diffusive concentration gradients, and their effect on monocytes transendothelial migration will help us to have a better understanding of underlying mechanisms during an inflammatory response in early stages of atherosclerosis. The results of this research will lead to the development of an improved *in vitro* model to test therapeutic strategies associated with inflammatory diseases.

## 2. Methods

*2.1 Cell culture*

Human aortic endothelial cells (HAEC) and endothelial cell complete medium were purchased from PromoCell, Heidelberg, Germany. HAEC were cultured in 75 cm$^2$ flasks and incubated at standard conditions (37 °C and humidified atmosphere of 5% $CO_2$) and used at 90% confluence.

*2.2 Preparation of the 3D in vitro vascular tissue model*

Costar (Corning life sciences, Cambridge, MA) and ThinCert Transwells® (Greiner Bio-one, Frickenhausen, Germany) with 8.0 μm permeable membranes were used to create the 3D tissue model in 24 and 12-well inserts, respectively. The inserts provide easy access to the free factors released in the apical and basal compartments of the endothelial layer in the 3D tissue model for taking measurements that makes them more advantageous than other models. Collagen type I solution of the 3D tissue model was prepared using previously described protocols [54-56] and added to the top of the membrane (78 and 266 μl per well for 24 and 12-well inserts, respectively). Plates were incubated at standard conditions to form a gel. Complete medium was added to the upper and lower chambers and incubated overnight to equilibrate the collagen matrices. The next day, the gel was coated with fibronectin (Biomedical Technologies, Stoughton, MA) and seeded with HAEC ($7.5*10^4$ cells/cm$^2$). The cells were incubated and used for experiments at one day after reaching confluency and forming an endothelium by visual observation (Fig. 1).

To construct the collagen-coated and 2D models, membranes (8.0 μm pore size) were coated with collagen (15 μl per well for 24-well format inserts) and fibronectin; or fibronectin only, respectively, seeded with HAEC as stated above, and used at one day post-confluence.

*2.3 HAEC viability and trans-endothelial electric resistance*

Viability and unit area resistance of HAEC layer were measured under normal and activated conditions. To activate the HAEC, 10 ng/ml recombinant human TNF-α (R&D Systems, Minneapolis, MN) in complete medium was added to the apical surface of the HAEC, and the cells were analyzed at 1 and 24 h incubation. These times were selected in order to study cells behavior at both early and late time points. Samples treated for 24 h with Triton X-100 (1%, v/v) (Fisher Scientific, Pittsburgh, PA) were used as a negative control (NC) for cell death, and samples in complete media was used as a positive control (PC).

Viability was measured using CellTiter-Blue® assay (Promega, Madison, WI) following the manufacturer's protocol. This assay measures cellular metabolism by measuring the conversion of the dye resazurin to the fluorescent product resorufin, and only viable cells are able to reduce the former to the latter. Fluorescent intensities of the samples were measured using a fluorescent microplate reader (SpectraMax Gemini XS, Molecular Devices, Sunnyvale, CA) at excitation and emission spectra of 560 and 590 nm, respectively, and reported as relative fluorescent unit (RFU). Unit area resistance of the HAEC cultured in the models was also determined by measuring the trans-endothelial electric resistance (TEER) (WPI, Sarasota, FL) following the manufacturer's protocol. Average relative fluorescent intensities and unit area resistance of the HAEC obtained under normal and activated conditions were subtracted from the negative control values in order to obtain the true cell layer viability and resistance.

Furthermore, HAEC grown on collagen coated models were also used as control samples for the endothelium cultured on collagen matrices to investigate if the collagen per se could affect the cell behavior significantly. Average unit area resistance of the confluent HAEC grown on models were reported under normal conditions and subtracted from the unit area resistance of the cell free samples considered as background controls (BG).

*2.4 HAEC morphology*

HAEC were stained with media containing fluorochrome-conjugated PECAM (5 µg/ml) or isotype antibodies (Biolegend, San Diego, CA) according to standard cell staining protocol (BD Bioscience, San Jose, CA). The stained samples were removed from the inserts, mounted on slides, and the cells nuclei were counterstained with DAPI using ProLong® Gold antifade reagent (Invitrogen, Carlsbad, CA). Samples were examined by confocal microscopy to detect the cells nuclei and membrane. Images were analyzed using ImageJ software (version 1.49t) to determine the membrane surface areas of fifty cells per three independent samples.

*2.5 Expression of CAMs and MCP-1 on the endothelial cell surface*

The surface expression levels of VCAM-1, ICAM-1, PECAM-1, and MCP-1 were measured under normal and activated conditions. Cells were activated by adding TNF-α to the apical side of the endothelium, as mentioned before, and samples were analyzed after 1, 6, 12, 18, and 24 h incubation time points for the expression of CAMs and MCP-1. HAEC cell culture models incubated without TNF-α served as controls. For flow cytometry, cells were detached from the 2D model using human primary endothelial cell detach kit containing solution of 0.25% trypsin with EDTA (PromoCell), and membrane inserts were washed to collect the lifted cells. To remove the HAEC from the 3D tissue model, the collagen matrices were digested with collagenase D (Roche Applied Science, Indianapolis, IN). Briefly, collagenase D was added on the collagen matrices and incubated at culture conditions for 1-2 h to digest the matrices and the wells were washed to collect all detached cells. Cell suspensions were centrifuged (5 min, 1200 rpm, 4 °C), and immunofluorescence staining was done using anti-human fluorochrome-conjugated CD106 (VCAM-1), CD54 (ICAM-1), CD31 (PECAM-1), MCP-1 or isotype controls (all from Biolegend, San Diego, CA). A flow cytometer (FACSCalibur) was used to analyze the cells within a day of staining and data analyses were performed using FCS Express software (version 3). For each sample, uniform population of the endothelial cells were gated using side versus forward scatter plots to exclude dead cells and debris. The Overton subtraction method was used to determine percentage positive (PP) and net mean fluorescent intensity (MFI) of the cells population compared to the isotype controls for each marker [57]. The PP and MFI results of the activated HAEC were normalized to the values obtained for the control samples (media with no TNF-α stimulation).

*2.6 Release of MCP-1 within the 3D tissue model*

The kinetics of MCP-1 release from the HAEC grown in the 2D and 3D models in 12-well inserts were measured under normal and activated conditions. Cells were activated with TNF-α (as described previously), and MCP-1 release was quantified at 1, 6, 12, 18, and 24 h following addition of TNF-α. Culture models incubated without TNF-α were used as controls. To measure MCP-1 release from the 2D and 3D models, media from both the apical and the basal compartments of the endothelium were collected and stored separately at -20 °C. To quantify MCP-1 concentration in the matrix, the collagen matrix was digested by collagenase D (as described before), centrifuged, and the cell-free supernatants were stored at -20 °C. MCP-1 concentration was measured in the collected solutions using a human MCP-1 ELISA kit (BD Biosciences) following the manufacturer's protocol. Then, based on the collagen matrix volume calculations (manuscript in review) and the measured MCP-1 amount in the digested collagen solutions, MCP-1 concentration in the collagen matrices were determined.

*2.7 Preparation of monocytes*

Human peripheral blood mononuclear cells (PBMCs) were isolated from blood obtained from healthy donors (Oklahoma Blood Institute, Oklahoma City, OK), using a Ficoll-Plaque (GE Healthcare, Uppsala, Sweden) density gradient centrifugation method. Monocytes were negatively selected from the PBMCs using a pan-monocyte isolation kit (Miltenyi Biotech, Auburn, CA) using the manufacturer's protocol, and used immediately for adhesion and migration experiments.

*2.8 Monocytes adhesion and transendothelial migration*

Adhesion and migration of monocytes under normal and activated conditions were studied in both the 2D and 3D models in 12-well inserts. Activation was achieved by adding TNF-α to the HAEC apical layer, and samples were analyzed after 1 and 24 h. At the end of each incubation, culture media were removed from the endothelium apical compartment and human monocytes ($1.5*10^5$ cells/cm$^2$ in 0.5 ml complete medium) were added. After 2 h, samples were rinsed gently to remove the non-adherent or loosely adherent monocytes, and the entire cell population (HAEC and monocytes) were collected. Any monocytes that migrated to the endothelial basal layer were also collected. Monocytes that either adhered to or transmigrated across the endothelium were detected by immunofluorescence staining with anti-human CD14 fluorochrome-conjugated antibody or isotype control (BD Biosciences). Analysis was performed within a day of staining. Monocytes population was gated and the Overton subtraction method was used to determine the percentage of CD14 positive cells in each gate compared to the isotype controls [57]. Number of monocytes in gates was calculated based on the total number of events and the percentage of CD14$^+$ cells in each gate. Then, total number of monocytes in each sample was determined using the number of monocytes in each gate, instrument flow rate, time elapsed for collecting the events, and total initial sample volume.

*2.9 The effect of MCP-1 on monocytes adhesion and migration*

The effect of MCP-1 on monocytes adhesion and migration in the 3D tissue model was also studied using an MCP-1 neutralizing antibody. At the end of 1 and 24 h incubation time point with TNF-α, media was removed from the HAEC apical compartment and was replaced with fresh media containing 750 ng/ml purified anti-human MCP-1 antibody (Biolegend). The samples were incubated for 2 h, the media was removed, and monocytes were added to the apical side of the endothelium. After 2 h incubation, cells were collected and analyzed, as described previously.

*2.10 Statistical analysis*

All experiments were performed in replicate to determine the mean ± standard deviation (SD) of the samples. A nonparametric statistical analysis of Kolmogorov-Smirnov (K-S) test was used to determine if the distribution of cells sizes are equal between independent samples group. Student's t-test was used to determine the significantly different groups among pairs. *p* value of < 0.05 was considered significant.

## 3. Results

*3.1 HAEC viability and trans-endothelial electrical resistance*

Viability and unit area resistance of the HAEC layer, quantified for both models under normal condition and upon activation by TNF-α for 1 and 24 h, are shown in Fig. 2. To compare the viability of the HAEC, we measured the metabolic activity as an indicator of viability in the cells grown in the models using the Cell-Titer Blue® Assay. No significant difference in cell viability is observed for 2D and 3D models with the addition of TNF-α compared to positive control (PC) samples (Fig. 2A). Furthermore, HAEC viability is lower in the 3D tissue model compared to the 2D model, when cells are activated at early and late time points, but not statistically significant. Our results clearly show that the cells in the novel, 3D tissue model retain viable.

Resistance results, shown in Fig. 2B, indicate that cell layer integrity at normal condition does not alter significantly within 1 h stimulation in 2D and 3D models. However, significant 1.6 and 4.5-fold cell resistance reduction ($p < 0.05$) is observed after 24 h of stimulation compared to the positive control (PC) samples in the 2D and 3D models, respectively. TEER measurements results also shows a significant 1.9, 2.2, and 5.2-fold decrease ($p < 0.05$) in cell resistance for the 3D tissue model compared to the 2D model for positive control samples, and after 1 and 24 h adding TNF-α, respectively. These results suggest that using collagen matrix as the ECM substrate has significant effect on formation of cell-cell junctions and HAEC spreading behavior.

The foregoing observations led us to further our understanding of the effect of collagen material per se as opposed to the collagen matrix ECM on HAEC layer unit area resistance. Fig. 3 shows the unit area resistance of the HAEC layer grown on the microporous membrane, collagen coating, and collagen matrix models, quantified by measuring the TEER under normal condition subtracted from the background (BG) unit area resistance. TEER measurements show a significant 2.1 and 1.6-fold decrease ($p < 0.05$) in cell resistance in the collagen matrix model compared to the microporous membrane and collagen coated samples, respectively. The average unit area resistance of the cells is decreasing as the collagen volume is increasing in the coated and matrix models compared to the 2D model. Therefore, confluent HAEC layer grown in the 3D tissue model had the highest permeability under normal condition.

The results suggest that endothelium-collagen interaction per se has an influence on cell-cell junctional zones which has decreased the HAEC resistance. Moreover, it is shown that formation of cell-cell junctions pattern not only depends on the interaction of cells and the type of material itself used as the substrate, but also the endothelium resistance varies with using the substrate as subendothelium ECM as opposed to a single layer coating film.

*3.2 HAEC morphology*

In this section, we sought to investigate the influence of subendothelium substrate and ECM on HAEC morphology which led to the observed variations in HAEC unit area resistance in section 3.1. Representative images of HAEC cultured on the membrane, collagen coating, and collagen matrices are shown in Fig. 4A. From microscopy analysis, it is observed that HAEC morphology was similar but not identical on the microporous membrane, collagen coated, and the matrices. Generally, HAEC were circular and round shape more like cobblestones and in some cases appeared more elongated in different models. However, quantitative analysis of HAEC sizes in Fig. 4B shows physical variations of cells surface area distribution depending on the interaction between cells and their surroundings substrate.

HAEC grown on the membrane environments exhibit a broad diameter range (22-76 µm) compared to the collagen coated models whereas the diameter range is narrowed down (24-73 µm). However, based on the K-S statistical analysis, HAEC on the collagen coated samples have the same size distribution function as the cells grown on the membrane model. Furthermore, sizes of HAEC grown on the matrices appeared to be more uniformly developed with a smaller diameter range (20-53 µm) and a significant difference ($p < 0.01$) in distribution function compared to the HAEC grown on the membrane and collagen coated models. The significant junctional zone distribution difference of HAEC grown on the collagen matrices compared to the microporous membrane and collagen coated samples suggests that the endothelium unit area resistance ought to be different between the samples which matches the data shown in Fig. 3. These data suggest that not only collagen material itself had an effect on biophysical morphology behavior of HAEC, but also having an ECM under an endothelium altered cells growth and spreading properties.

Furthermore, HAEC packing density was determined according to the microscopy images and reported as 66266 ± 15265, 63252 ± 7468, and 67798 ±12713 cells/cm2 for microporous membrane, collagen coating, and collagen matrix models, respectively. Although HAEC size distribution pattern is different on the microporous membrane, collagen coating, and collagen matrix microenvironments, the average cells packing density is not changing significantly. Hence, we expected to get the same cell viability results in different models which correlates to the viability data presented in Fig. 2A.

*3.3 Expression of CAMs and MCP-1 on the endothelial cell surface*

The kinetic and magnitude of CAMs and MCP-1 expression on the HAEC surface was measured for control and TNF-α activated samples by direct immunofluorescence staining and detection by flow cytometry. Results are presented in Fig. 5 as ratio of the net mean fluorescent intensity (MFI) and percentage positive (PP) of cell population in activated samples to the control (media with no TNF-α stimulation).

There was a low expression of VCAM-1 on unstimulated endothelial cells (data not shown), and a significant upregulation of VCAM-1 expression after 24 h TNF-α treatment (15.4-fold increase compared to control). The PP of the cells population expressing VCAM-1 is not changing significantly after 6 h activation for both models. However, the MFI of the VCAM expression is increasing gradually between 6-18 h when it increases significantly after that up to 24 h for both models. At 24 h stimulation by TNF-α, the MFI of VCAM-1 positive cells in the 2D model is significantly higher than cells in the 3D tissue model (2.3-fold, $p < 0.01$).

ICAM-1 is expressed on normal endothelium in control samples (data not shown), but the expression is upregulated after stimulation by TNF-α with a time course similar to that of VCAM-1. Increase of ICAM-1 membrane expression PP reached a plateau after 6 h TNF-α activation. However, the MFI of ICAM membrane expression is increasing in both models. ICAM-1 membrane protein expression was detected to a lesser extent in the 3D tissue model than the 2D model at both early and late time points. It was shown that the MFI expression of ICAM-1 protein were statistically significantly higher in HAEC treated with 24 h TNF-α in 2D models compared to the 3D tissue models (1.9-fold, $p < 0.05$). It is also worth noting that the increase in PP cells expressing VCAM-1 after activation is significantly higher than the positive cells for ICAM-1 membrane protein compared to the control samples.

PECAM-1 expression localized at cell–cell borders of HAEC confluent monolayers is not upregulated in response to TNF-α compared to the control samples without stimulation. No significant differences of PECAM-1 expression (MFI and PP) were observed on the cells surface between the 2D and 3D models which shows that TNF-α did not have effect on PECAM expression in both models over 24 h activation.

The expression of MCP-1 was observed on the cells surface with not a significant difference between the 2D and 3D models after activation. According to the flow cytometry results, moderate number of HAEC population has expressed MCP-1 after 6 h activation (16.1-fold increase in PP), but with a weak intensity (1.4-fold increase in MFI). The increase of MCP-1 membrane protein expression (both PP and MFI) reached a plateau after 6 TNF-α incubation when it remains unchanged till 24 h stimulation.

The data suggest that TNF-α increases the level of VCAM-1 and ICAM expression on the HAEC surface in both models. And the HAEC from the 2D models have a higher expression of the key CAMs intensity (MFI) and the same expression of MCP-1 associated with monocytes cell migration than the HAEC from the 3D tissue model after 24 h stimulation.

*3.4 Release of MCP-1 within the 3D tissue model*

The kinetic of MCP-1 release from the HAEC over 24 h time point in both the apical and basal side of the HAEC for the 2D and 3D models was measured by ELISA and shown in Fig. 6. The data demonstrate that HAEC had the highest release of MCP-1 in the apical and basal compartments of endothelium after 24 h TNF-α stimulation in both models with a significant difference between the 2D and 3D models at each activation incubation time point.

Although the final concentration of MCP-1 in the endothelial basal layer is shown to be identical in the 2D and 3D models, in fact a significant difference in the net release of MCP-1 is observed in the collagen matrix of the 3D tissue model compared to the HAEC basal layer in 2D model. After 24 h activation, HAEC released MCP-1 with 2.9 and 11.2–fold increase compared to the corresponding control samples in the 2D and 3D models, respectively. Accordingly, due to the sample concentrated in the collagen matrix, resulting in an abrogated release into the surrounding media, an earlier detection of MCP-1 release was expected from the 3D tissue model. As it is shown in Fig. 6, there is no significant difference in the MCP-1 concentration released to the basal layer of HAEC in the 2D model after 6 h TNF-α activation compared to the control sample. However, stimulating HAEC in the 3D tissue model for 6 h had a significant effect on MCP-1 release in the collagen matrix beneath the HAEC layer (9.5-fold, $p < 0.01$).

Furthermore, the results display that the ratio of total MCP-1 mass released when HAEC were activated to the control samples increased significantly as the TNF-α incubation time was increasing in both models. Moreover, the normalized MCP-1 amount released is exhibited to be statistically significantly higher in the 3D tissue model compared to the 2D model starting at 6 h stimulation up to 24 h when the ratio was 1.6-fold higher in the 3D tissue model than 2D model ($p < 0.05$). Taken together, these results indicate that having the collagen matrix beneath the endothelium in the 3D tissue model had altered the cells behavior in producing MCP-1 before and after activation which will have an influence on monocytes migration later on.

*3.5 Monocytes adhesion and transendothelial migration*

Total number of monocytes that adhered and migrated across the endothelial cell layer for each model was determined by direct immunofluorescence staining and detection by flow cytometry. The percentage of monocytes that adhered to and migrated through the endothelium after 2 h is presented in Fig. 7. The results show that there are significantly more monocytes transmigration occurred in the 3D tissue model compared to the 2D model after 24 h of stimulation (2.2-fold, $p < 0.1$) and the negative control (NC) samples without TNF-α treatment (2.4-fold, $p < 0.1$), which shows the effect of different MCP-1 concentration gradient profile on monocytes migration. An increase in the number of monocytes that migrate across the endothelial layer for the 3D tissue model compared to the 2D model, indicates that MCP-1 influence is greater in the 3D tissue model, due to the MCP-1 haptotactic gradient. The greater number of monocytes for the HAEC in the 3D tissue model can be attributed to the difference in MCP-1 within the collagen of the 3D tissue model compared to the free MCP-1 in the 2D model.

We also studied the specificity of the MCP-1 gradient on the monocyte migration by neutralizing the available MCP-1 with MCP-1 antibody. The results show that there is a significant reduction in the number of monocytes adhered and transmigrated between the anti-MCP-1 antibody-treated samples versus the 24 h activated samples (7.2-fold, $p < 0.05$). Our results clearly show that treatment with the MCP-1 antibody could significantly reduce MCP-1 induced monocytes migration through the collagen matrices.

## 4. Discussion

In this study, an advanced 3D *in vitro* vascular tissue model was introduced as a novel tool to study the early cellular mechanisms involved in atherosclerosis plaque formation. Using this new model, the effect of MCP-1 concentration profile on monocytes migration was determined. The results supported our hypothesis that MCP-1 transport was different in the 3D model than the traditional 2D microporous membrane model, which resulted in a difference in monocytes transendothelial migration between the two models.

Expressed chemokines such as MCP-1 diffuses into the ECM where they can bind to the ECM proteins and maintain, so ECM proteins are important factors for immobilizing chemokines and preserving the profile [15, 54]. Despite of this fact, the nature of the MCP-1 gradient and its effect on monocytes migration has been very difficult to assess in an *in vivo* setting and its mechanism is still not clear and is a very controversial area.

2D cell culture is widely used to investigate the MCP-1 concentration gradient effect on monocytes migration. Endothelial cells are grown on a thin, microporous membrane, the MCP-1 is added to the aqueous media below the membrane, and monocytes are added to the apical surface of the endothelial layer. MCP-1 is added to both the apical and basal side of endothelium layer for chemotactic control [41]. Monocytes moved across the pores toward the MCP-1 solution and transmigration of the cells was observed [42]. The number of transmigrated monocytes across the endothelial cell layer and the microporous membrane is counted for different concentrations of MCP-1 and different time points [58]. This model may not adequately predict *in vivo* cell behavior due to the lack of the third dimension which is an important factor in many physiological conditions. Interaction between cells and the below matrix and also the cell migration microenvironment condition are the most important issues that have to be considered in cell culture studies which are not achievable in 2D models. In order to address these issues, subsequent studies have proposed some alternatives as a development of 3D tissue models in laboratories [59, 60]. In these studies, 3D *in vitro* vascular tissue model is introduced as a tool to study monocytes transendothelial migration within the ECM. 3D tissue model consisting of a type I collagen matrix would be a better experimental model to mimic the subendothelial Human Endothelial Cells (HEC) as it is a major constituent of many tissues that make it useful for many model applications, and it can be obtained in a pure form without mixture of other bioactive ECM proteins which will simplify the experiment and the model [53, 54]. It is more likely to have ECM below the endothelial layer, so the secreted chemokines such as MCP-1 and its corresponding profile can concentrated as a fluid phase or bound to the beneath matrix compound [54, 59]. The 3D tissue model provides the added dimension that is important for the creation of a diffusive concentration gradient formed in the ECM, which is responsible for the control of many cellular mechanisms [54]. Such a 3D tissue model can be used *in vitro* cellular model to construct cell-immunity model [54, 60]. After passing across the endothelial layer monocytes can continue migration via chemoattractants [60] known as MCP-1 [16, 54] in the beneath space.

In this study, we showed that there were significant differences in HAEC response to TNF-α and monocytes transendothelial migration between the 2D and 3D models. Significant differences in HAEC response included the expression of CAMs and the permeability in response to TNF-α stimulation. The HAEC from the 3D model had a greater and earlier expression of the key CAMs associated with monocytes cell migration than the HAEC from the 2D model. Although the HAEC in the 2D model showed lower permeability overall compared to the 3D model, the cell layer showed a much higher increase in permeability after

TNF-α stimulation as measured by both TEER measurements and PECAM-1 expression. The greater expression of MCP-1 for the HAEC in the 3D model can be attributed to the difference in MCP-1 within the collagen of the 3D model compared to the soluble MCP-1 in the 2D model. This haptotactic gradient cannot be created within a 2D model. Significant differences in monocytes transendothelial migration between the 2D and 3D model showed that overall there is greater transendothelial monocytes migration in the 2D model compared to the 3D model; however, there are more MCP-1$^+$ cells that migrate in the 3D model. The greater number of cells migrating in the 2D model is a result of the greater permeability for the HAEC in response to TNF-α for the 2D model than for the 3D model. An increase in the number of MCP-1$^+$ monocytes that migrate across the endothelial layer for the 3D model compared to the 2D indicates that MCP-1 influence is greater in the 3D model, possibly due to the haptotactic influence, and that other factors may be driving cell migration in the 2D model.

The 3D model was compared to existing 2D models in order to show the influence of the ECM making up the third dimension on cell behavior. The expression of CAMs on the endothelial cells, MCP-1 release, MCP-1 diffusive gradients, and monocytes migration were compared between the 3D and 2D models. Findings from comparing the two models contribute to a better understanding of how cells behave in 2D versus 3D models and the development of improved experimental models. Proposed mechanisms can be used to target the MCP-1 involved during the inflammation with highly specific therapeutic strategies. A long term goal of our research group is utilizing the advanced 3D *in vitro* model to test preventative and therapeutic interventions of atherosclerosis.

Overall, the 3D *in vitro* tissue model is utilized in this project to characterize cellular adhesion molecules (CAMs) and to target one specific chemokine marker (MCP-1) that are critical participants in the transmigration of monocytes involved during the formation of the atherosclerosis lesion. The effect of MCP-1 local concentration gradients on monocytes migration is investigated using the 3D model. HAEC behavior (the CAMs and the chemokine expression) and transport properties (MCP-1 release profile) is also studied between the 2D cell culture and the 3D model. Differences between two models characteristics, such as MCP-1 concentration gradients and cell behavior, and their effect on monocytes migration, i.e., the effect of diffusive gradients in a 3D environment compared to a 2D model, will help us to have a better understanding of underlying mechanisms in atherosclerosis.

In conclusion, we have shown that kinetics and the level of CAMs expression in the 2D model is different from the 3D tissue model which has an effect on monocytes adhesion and transmigration through the endothelial layer between the two models. This 3D tissue model consisting of a collagen matrix would be a better alternative experimental model to mimic the subendothelial ECM. Such a 3D tissue model provides the added dimension that is crucial for the creation of diffusive concentration gradients which is an important factor in many physiological conditions. So this 3D model can be used as an effective tool for studying the monocytes transendothelial migration through the endothelial layer. The concentration gradient formed in the 3D model is distinctly different compared to the one in the 2D cell culture model, where the secreted factors from the endothelial cells dissolve quickly into the surrounding media. Furthermore, the 3D model provides a highly controllable micro-environment for investigating cellular interactions and responses [53]. Apart from this advantage, when focusing on the transmigration of monocytes, this matrix can also provide an area for monocytes to localize and differentiate into macrophages after the transendothelial migration [56]. This latter property of the 3D tissue model makes it applicable as a testing device for many diseases like atherosclerosis.

Many questions still remain about the expression of MCP-1 from endothelial layer and its concentration profile in subendothelial tissue. More studies in this area will lead to the development of an improved *in vitro* model that exhibits native characteristics of *in vivo* in order to study transendothelial monocytes migration associated with MCP-1 profile and inflammatory conditions. Furthermore, as the next phase of this research, we expect utilizing the developed 3D *in vitro* tissue model to develop a second generation mathematical model that describes MCP-1 transport within the 3D model, which will include a source term of MCP-1 production from the HAEC in response to TNF-α and the binding reaction of MCP-1 to the collagen matrix. Validating the mathematical model by comparing the obtained numerical results to the experimental measurements of MCP-1 in the 3D model and determining any correlation to monocytes migration would also be conducted. With the completion of ongoing future phases, an advanced *in vitro* tissue model will be developed and can be used to study inflammation in atherosclerosis-associated monocyte migration diseases.

**Acknowledgement**

This work was supported by a grant from the National Institute of Biomedical Imaging and Bioengineering (1R15EB009527-01).

**Figure Legends:**

**Figure 1.** The steps to construct the 3D tissue model within the Transwell permeable supports. Briefly, a 57.1 vol % bovine collagen type I solution was prepared and added to the top of the membrane and incubated at 37 °C and 5% $CO_2$ to form a gel. Complete endothelial cell growth medium was added to the top and bottom chamber and incubated overnight at standard conditions to equilibrate the collagen matrices. The next day, the matrices were seeded with HAEC ($7.5*10^4$ cells/cm$^2$) and cells confluency was monitored everyday by visual observation. Samples were used for experiments at post-day reaching confluency.

**Figure 2.** HAEC viability fluorescent intensity measurements (ex/em: 560/590 nm) **(A)** and unit area resistance **(B)** in the 2D and the 3D models under normal condition (positive control: PC) and when the models were activated with TNF-α (10 ng/ml) for 1 and 24 h. Values are presented as mean ± SD of absolute differences compared to the Triton X-100 (1%, v/v) treated samples (negative control: NC); n=3; * indicates $p$ value < 0.05.

**Figure 3.** HAEC unit area resistance in the microporous membrane, collagen coated, and collagen matrix models under normal condition. Values are presented as mean ± SD of absolute differences compared to the background (BG); n=3; * indicates $p$ value < 0.05.

**Figure 4.** HAEC were immunostained for the nuclei and endothelial marker PECAM-1. **(A)** Confocal microscopy was used to image the aortic endothelial cell shape in microporous membrane, collagen coated, and collagen matrix models. Nuclei and PECAM protein expression were revealed by DAPI staining (blue) and CD31 antibody (red), respectively. All the images were taken at the same magnification of 400× and at the same settings as their staining isotype control IgG. Scale bar, 50 µm. **(B)** Membrane surface areas of 50 cells per

sample were analyzed for three independent samples (n=3) using ImageJ software (version 1.49t). Values for each cell are shown in different models and the average surface area is marked with a solid line. ** indicates *p* value < 0.01.

**Figure 5.** HAEC membrane protein expression flow cytometry analysis of CAMs/MCP-1 in the 2D and 3D models. For each sample, uniform population of the endothelial cells were gated using side versus forward scatter plots to exclude dead cells and debris. The Overton subtraction method was used to determine percentage positive (PP) and net mean fluorescent intensity (MFI) of the cells population compared to the isotype controls for each marker. Results are presented as mean fluorescent intensity (MFI) and percentage positive (PP) of each marker in activated samples normalized to the control samples (media with no TNF-α stimulation); n=3, MFI of HAEC ICAM expression is significantly higher in 2D models compared to the 3D tissue models after 6 h activation with TNF-α (* *p* < 0.05).

**Figure 6.** MCP-1 concentration (ng/ml) released from HAEC in the apical and basal of endothelium in 2D and 3D models for TNF-α activated (10 ng/ml) and control (normal condition) samples at different time points. Arrows show concentration elevation from control to activated condition. And ratio of total MCP-1 mass (ng) released from HAEC after treating with TNF-α (10 ng/ml) to MCP-1 mass released in control samples (normal condition) in the 2D model compared to the 3D tissue model at different time points. Values are presented as mean ± SD; n=3, MCP-1 concentration is significantly higher in activated samples compared to control (* *p* < 0.05).

**Figure 7.** Monocyte adherence and migration across the HAEC layer for each model in response to MCP-1 gradients for control and activated samples. HAEC were treated with 10

ng/ml TNF-α for 1 and 24 h. TNF-α treated 2D and 3D models and 3D tissue models incubated with an excess of MCP-1 neutralizing antibody (neutralizing Ab) were compared. Samples incubated with complete media without TNF-α was used as a negative control (NC). Human monocytes ($1.5*10^5$ cells/cm$^2$ in 0.5 ml media) were added on the apical layer of HAEC and the samples were incubated at standard conditions for 2 h. At the end of the incubation time, the top surface was rinsed and the number of monocytes that adhered on or migrated through the endothelium was determined. Values are presented as mean ± SD of the percent of attached and migrated monocytes normalized to the initial number of monocytes added to each well; n=3; # and * indicate $p < 0.1$ and 0.05, respectively, for change in percentage of adhered and migrated monocytes.

**Figure 1.**

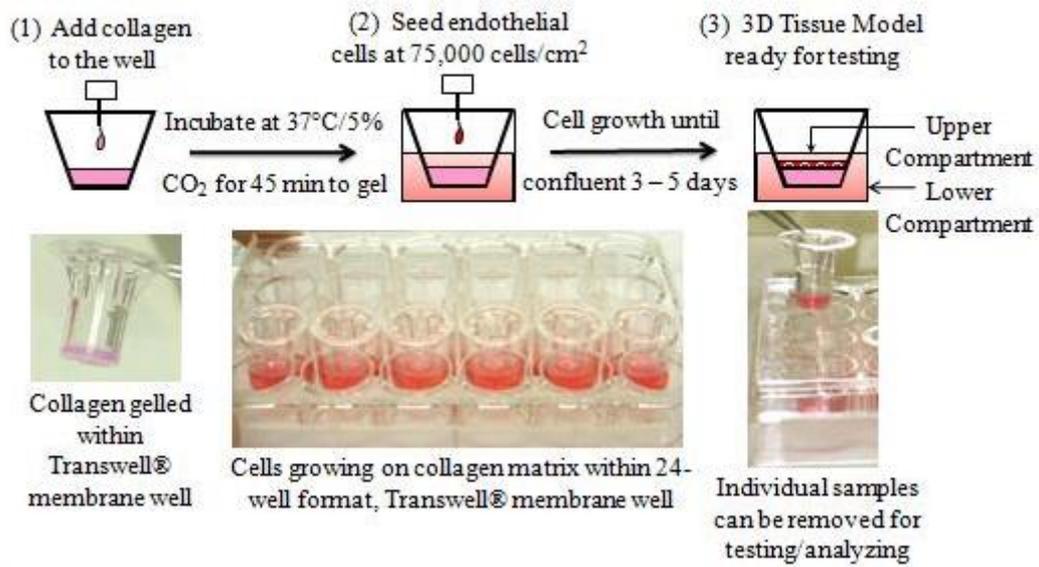

**Figure 2.**

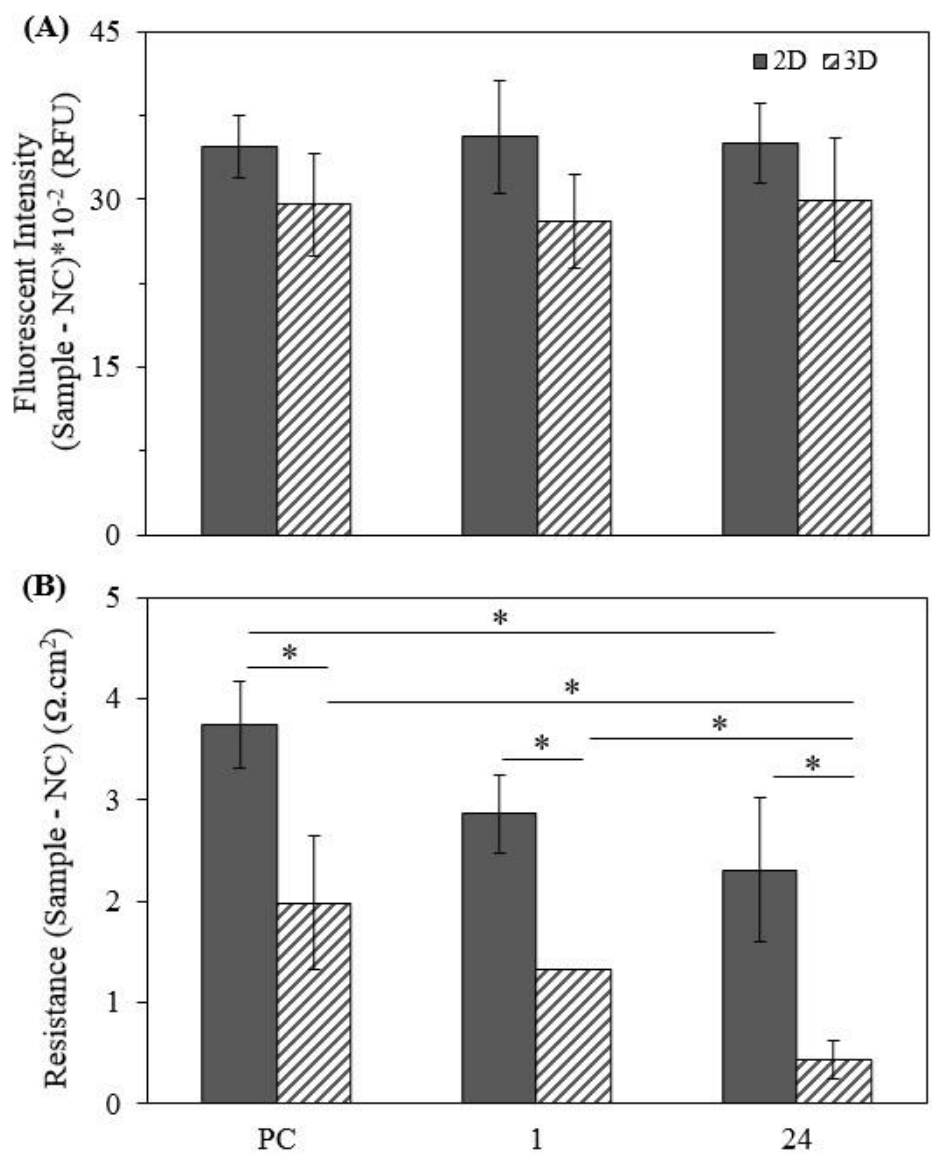

**Figure 3.**

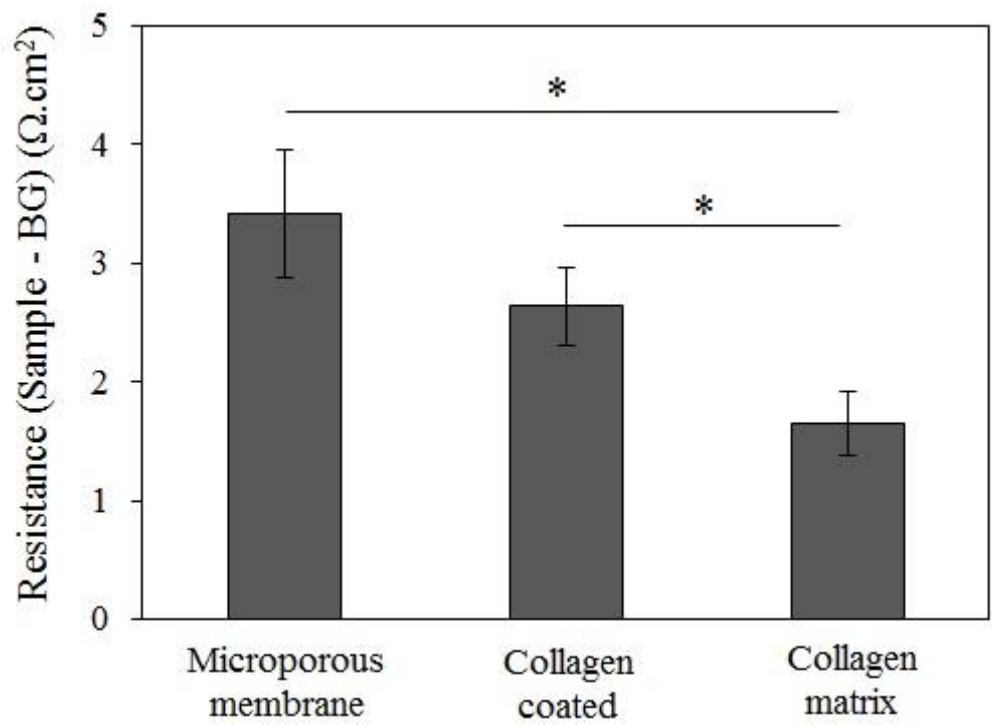

**Figure 4.**

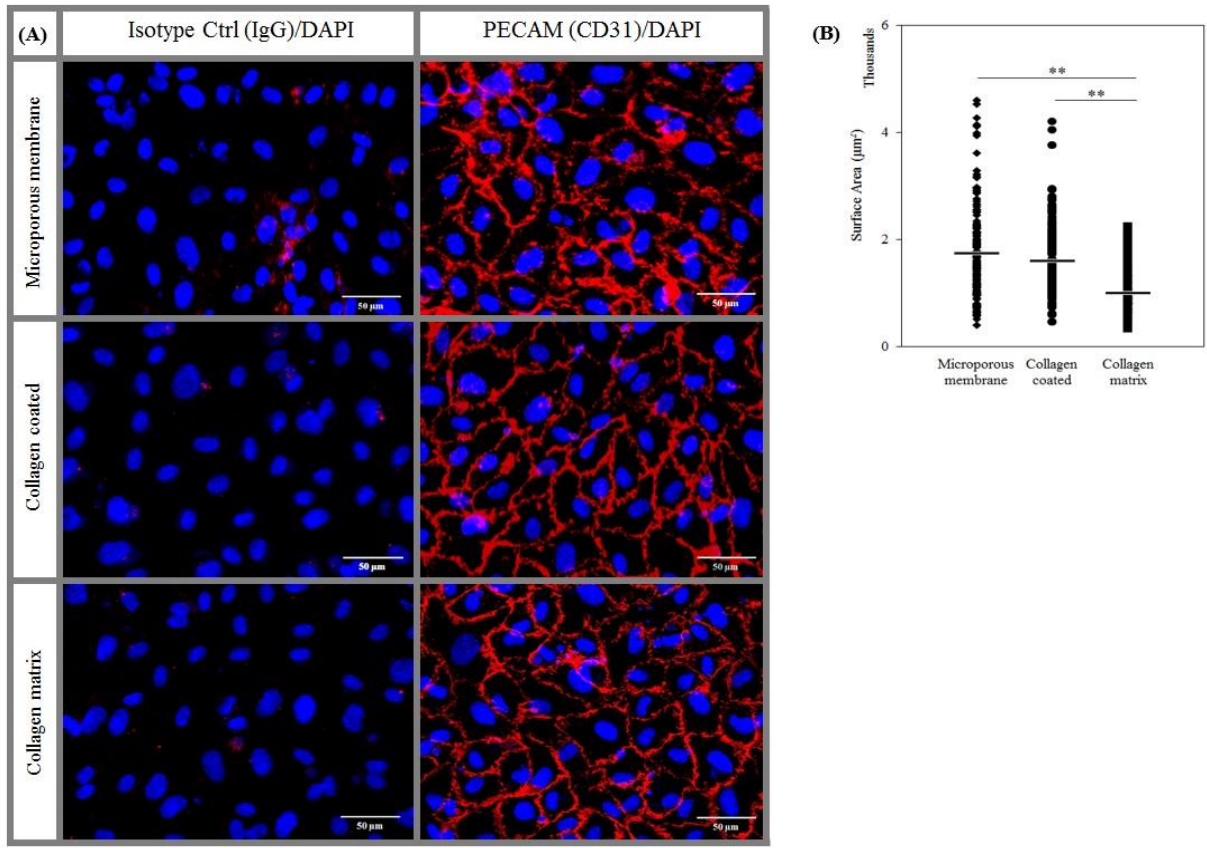

**Figure 5.**

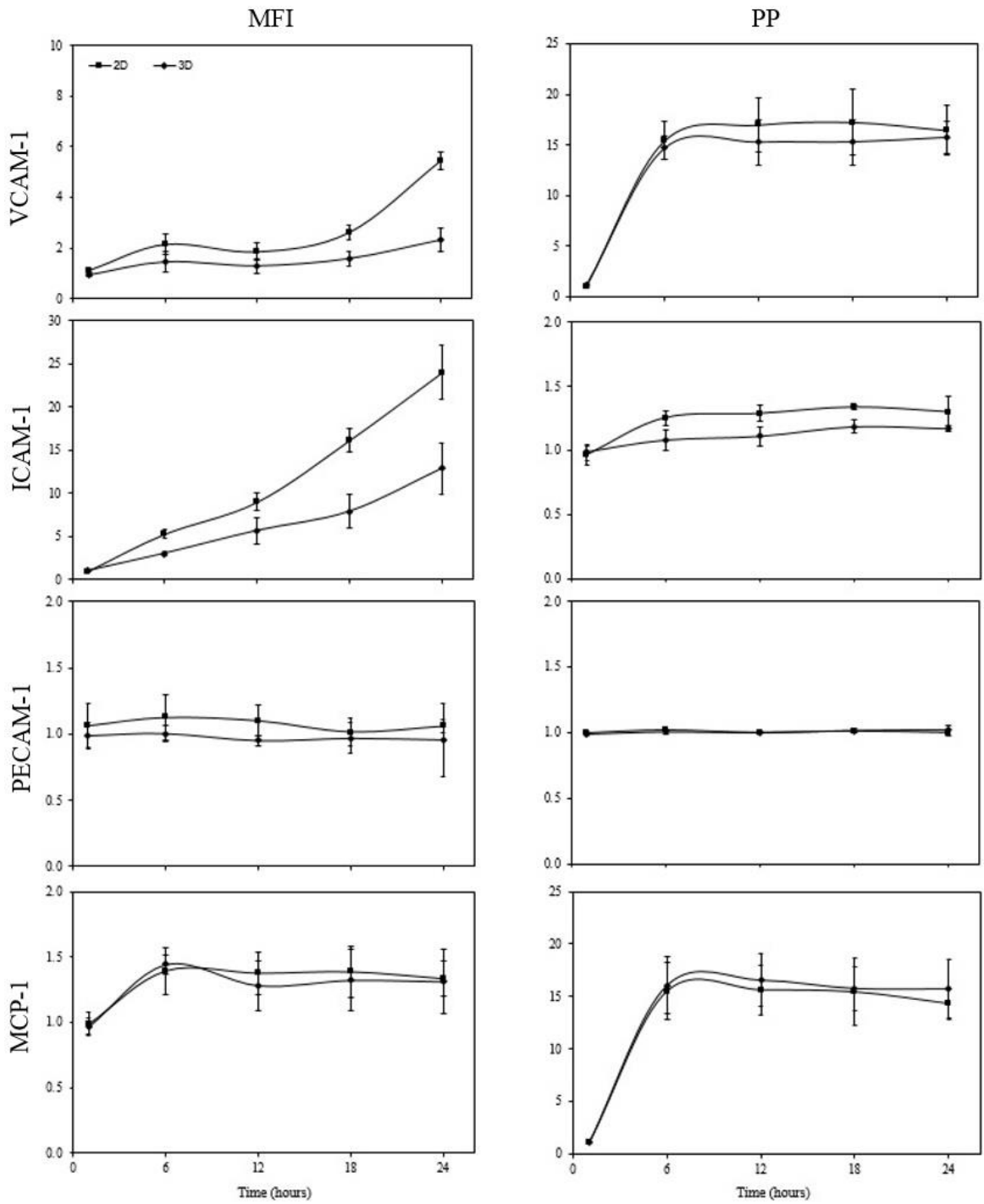

**Figure 6.**

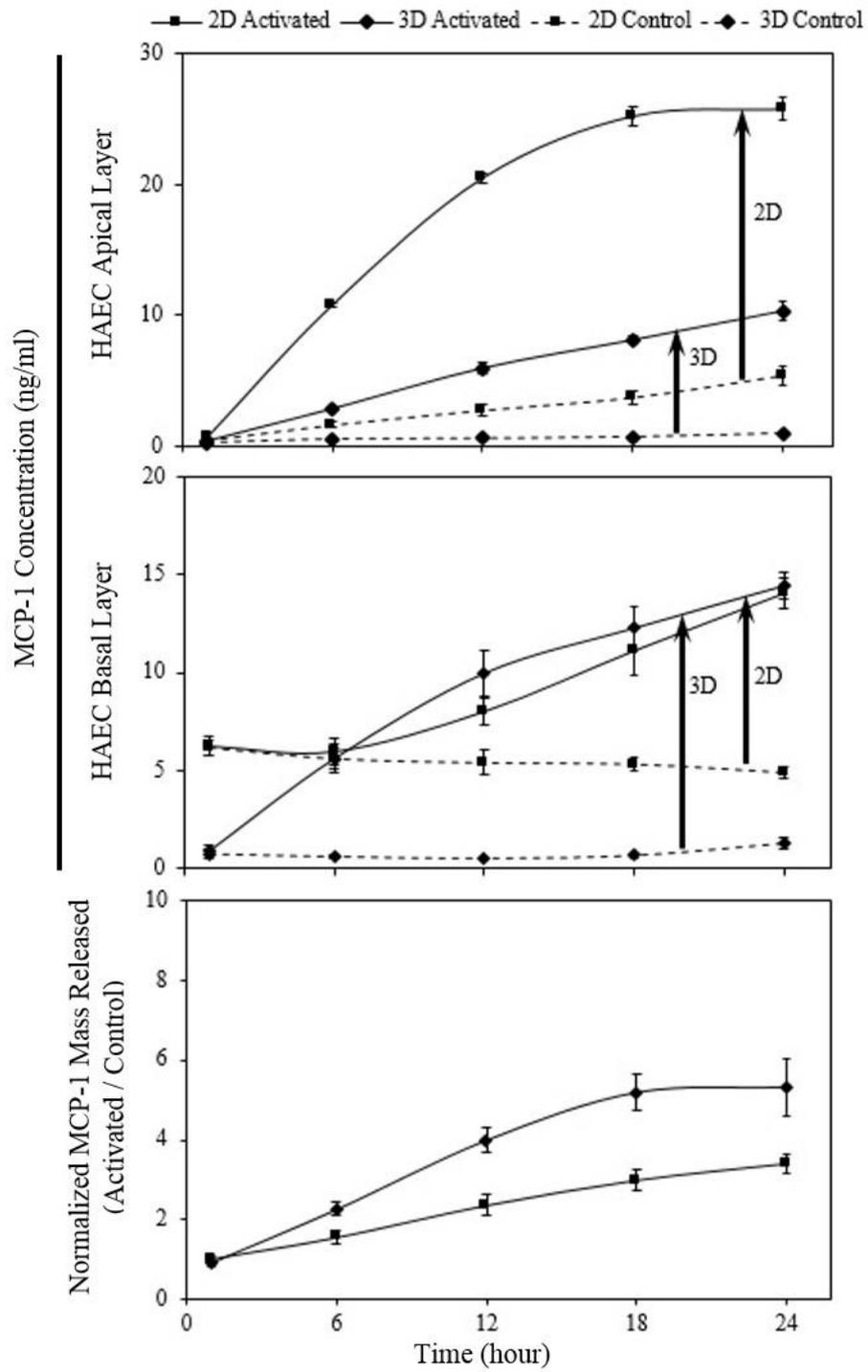

**Figure 7.**

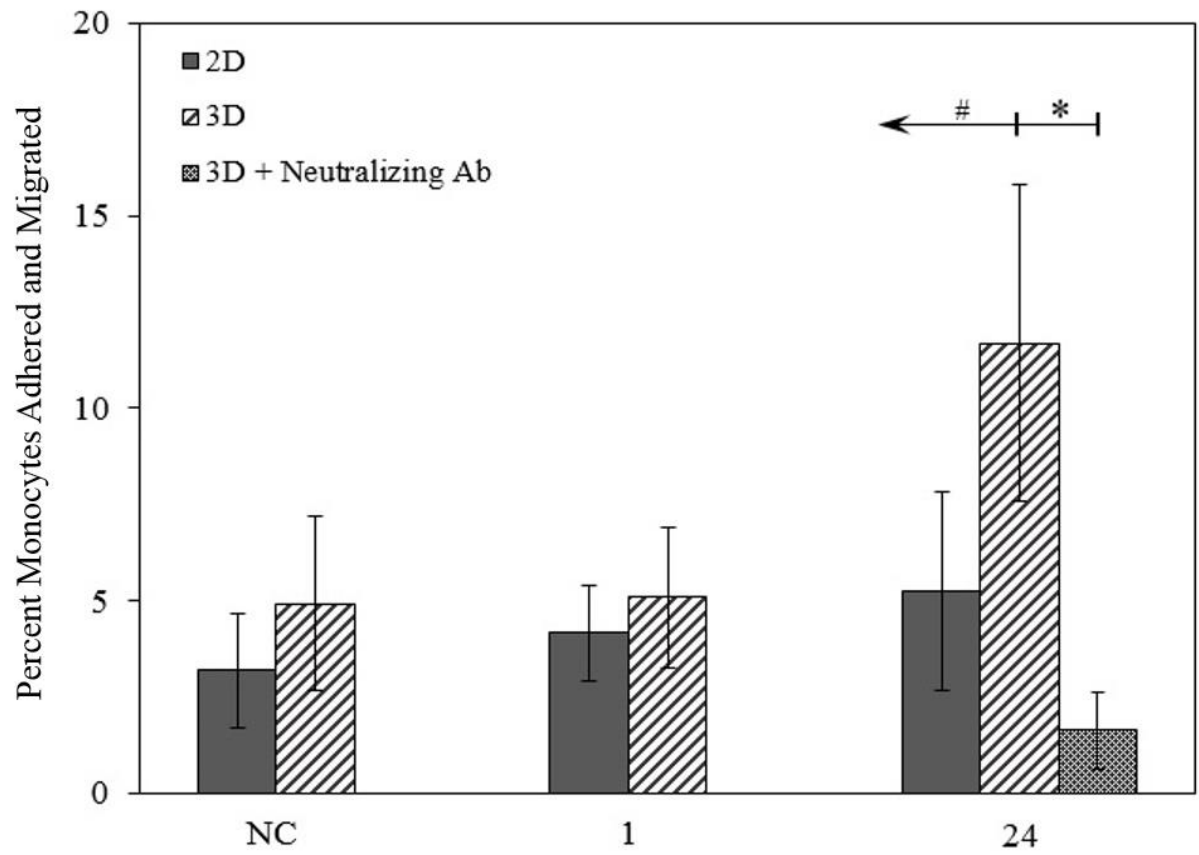